\documentclass[aps,prl,preprint]{revtex4}
\usepackage{graphicx}

\begin{document}

\title{Localization and Anomalous Transport in a 1-D Soft Boson Optical Lattice}

\author
{A. K. Tuchman, W. Li, H. Chien, S. Dettmer, M. A. Kasevich}
\affiliation{Physics Department, Stanford University, Stanford CA,
94305}

\date{\today}

\begin{abstract}
We study the dynamics of Bose-Einstein condensed atoms in a 1-D
optical lattice potential in a regime where the collective
(Josephson) tunneling energy is comparable with the on-site
interaction energy, and the number of particles per lattice site is
mesoscopically large. By directly imaging the motion of atoms in the
lattice, we observe an abrupt suppression of atom transport through
the array for a critical ratio of these energies, consistent with
quantum fluctuation induced localization. Directly below the onset
of localization, the frequency of the observed superfluid transport
can be explained by a phonon excitation but deviates substantially
from that predicted by the hydrodynamic/Gross-Pitaevskii equations.
\end{abstract}

\maketitle

Coherent control of the collective dynamics of macroscopic quantum
systems is currently of great interest due to possible applications
in quantum measurement and information science \cite{Bouwmeester,
Neilson}.  For example, coherent manipulation of superconducting
currents in Josephson junction circuits has led to the realization
of high-$Q$ electronic qubits for quantum logic devices
\cite{Devoret}.  Similarly, manipulation of the superfluid
properties in atomic systems, such as with BECs in optical lattices,
may soon provide a realization of de Broglie wave interferometers
which perform below the shot-noise limit \cite{Bouyer}.

At zero temperature, the physical characteristics of coupled
superconducting/superfluid reservoirs are determined by two
competing energies: the kinetic energy associated with tunneling
between sites ($E_J$), and the on-site interaction energy ($E_C$),
resulting from (repulsive) inter-particle interactions
\cite{Paraoanu, Burnett}.  Specifically, for a BEC in an optical
lattice system $E_J \equiv N\gamma$, where N is the number of atoms
in a lattice site and $\gamma$ is the inter-site tunneling energy,
and $E_C \equiv g\beta$, where $g\beta$ is the mean-field energy.
The nature of the many-body ground state in the lattice array is
governed by the ratio $ \Gamma \equiv {E_C}/{E_J} \equiv
{g\beta}/{N\gamma}$, which can be divided into three regimes. For
$\Gamma\ll1$, the system exhibits global superfluidity and
long-range phase order. As $\Gamma$ approaches $1$, interactions
lead to a frustration of long-range phase order and a corresponding
reduction of the single reservoir number variance $\delta N$, with
$\delta N \sim (N\gamma/g\beta)^{1/4}$. When $\Gamma \sim 1$, the
system undergoes a transition to an insulating regime where number
fluctuations are strongly suppressed ($\delta N < \,1$) for
commensurate filling. For translationally invariant lattice arrays,
this defines the Mott-insulating regime (MI) \cite{Fisher, Bloch2}.
This system has been shown to map onto 1-D superconducting chains,
which demonstrate a Kosterlitz-Thouless transition \cite{Doniach}.

Previously, interferometric techniques have been used to study the
ground state properties of BECs in optical lattices forming 1-D
arrays with high filling factor in the regime $\Gamma>1$
\cite{Orzel}. This work demonstrated that, for the system studied,
interferometric measurements do not have the specificity to reveal
possible abrupt changes in the many-body state of the system as the
Mott-insulating regime is reached \cite{Bloch}. However, recent
theoretical work has shown that transport measurements in the regime
$\Gamma\sim 1$ are expected to provide additional insights into the
superfluid properties of the array \cite{Roth}. Similar transport
measurements have previously been used to study the superfluid
properties of arrays in the semiclassical regime, $\Gamma\ll1$, and
have observed both superfluid transport \cite{Ingu} as well as
damping. Dissipation has been attributed primarily to either
dynamical instabilities \cite{Ingu2} or quantum fluctuation effects
\cite{Fertig, Esslinger}, depending on the experimental parameters.
Both of these mechanisms make it difficult to observe finite
amplitude transport for systems in their ground state at deep
lattice depths, even for $\Gamma<1$.

In this work, however, we observe residual coherent transport in
this dissipative regime, which we attribute to the presence of
anomalous phonon excitations.  We study this transport through
direct observation of atom motion across a 1-D lattice array
superimposed on a weaker harmonic potential.  We observe a crossover
from coherent oscillatory behavior to localization as we vary the
strength of the coupling between adjacent lattice sites.  We
attribute this localization to the role of quantum fluctuations in
driving the formation of strongly correlated, mesoscopic insulating
states with lattice site atom number occupation, $N \sim 100$. We
support this interpretation by observing the cessation of
macroscopic tunnelling to occur when $\Gamma \sim 1$, the same
critical relation found in analogous Josephson junction experiments
\cite{Mooij1}. Furthermore, we rule out localization due to known
semiclassical effects.

The Bose-Hubbard Hamiltonian is expected to provide an accurate
theoretical description of a lattice array of bosonic atoms
\cite{Jaksch}. Written in terms of the on-site single particle
creation and annihilation operators $\hat a_{i}^{\dagger}$ and $\hat
a_{i}$:
\begin{equation}
H = -\gamma{\sum_{<i,j>}}\hat a_{i}^{\dagger}\hat a_{j} +
{}\sum_{i}\frac{1}{2}g \beta_i\hat a_{i}^{\dagger}\hat
a_{i}^{\dagger}\hat a_{i}\hat a_{i} +\sum_{i}{V_i}\hat
a_{i}^{\dagger}\hat a_{i}
\end{equation}
where $V_i=\Omega i^2$ is the external potential due to the harmonic
magnetic trap, the subscript {\it{i}} denotes the {\it{i}} th
lattice site, and $g=4 \pi \hbar^2 a/m$ ($a$ is the repulsive s-wave
scattering length and $m$ is the atomic mass). $\beta_i$ and
$\gamma$ are determined from integrals over single particle
wavefunctions \cite{integrals}, and can be precisely, experimentally
controlled by raising and lowering the intensity of the optical
lattice in order to vary $\Gamma$.

In our experiments, we investigate the response of the lattice
ground state to sudden shifts in the position of the harmonic
potential used to initially create the condensate. Exact theoretical
predictions using Eq. 1 for the ensuing array dynamics are difficult
for our experimental conditions due to the large Hilbert space
needed to model the system. Nevertheless, we gain intuition into the
dynamic behavior by considering limiting regimes. For static
potentials and translationally invariant arrays, a second order
quantum phase transition from a superfluid to a MI phase is
predicted to occur at $\Gamma \sim 1$ \cite{Doniach,Fisher, Sachdev,
Monien}. At this point the many-body wavefunction localizes to a
product of states with nearly quantized atom number at each lattice
site for commensurate lattice filling. Such states lack macroscopic
phases, and thus, we expect suppression of Josephson-like transport
of atoms through the array when an external driving force is
applied. On the other hand, just below the transition, particle
fluctuations $1 < \delta N \ll N^{1/2}$ are sufficient to to define
an average macroscopic phase and induce superfluidity, even for
large filling factors.

Inclusion of the external harmonic potential (as in Eq. 1)
complicates the analysis \cite{Jaksch, Prokofev, Polkovnikov,
Troyer}. For deep lattices, where $\Gamma > 1$ across the entire
array, mean-field analysis predicts isolated Mott domains of
incommensurate filling with local incompressible regions but does
not demonstrate the vanishing global compressibility indicative of a
true phase transition \cite{Troyer}. These domains have been
recently observed in Refs. \cite{Ketterle,Gerbier}. For shallower
lattices, regions near the edges of the array can locally fulfill
the Mott-insulator condition, while the central lattice sites remain
in the superfluid regime.  We expect the formation of these
insulating barriers to substantially alter the dynamics of the
superfluid confined in the interior regions.

For a given lattice strength, $U$, we determine the values of $g
\beta_i$ and $N_i$ using the following two-step approach. First, we
numerically solve the 3-D Gross-Pitaevskii equation (GPE)
\cite{Pethick} for a single lattice site (neglecting tunneling
between adjacent lattice sites) to determine the spatial
wavefunction associated with the individual lattice sites as a
function of the strength of the lattice potential and the number of
atoms in the well.  This allows determination of the chemical
potential $\mu_i$ and effective value of $g \beta_i$ \cite{Zapata,
Stoofa}. Analytic expressions for these quantities can be obtained
in the limit where the kinetic energy is negligible. For our
parameters, the analytic expressions agree to within 10\% with
numerical estimates that do not neglect kinetic energy. Next, we
estimate the equilibrium distribution of the mean occupancy of each
lattice site across the array by equating the chemical potentials
associated with each site, subject to the constraint that the total
number of atoms sums to a fixed value. We obtain analytic
expressions if we neglect kinetic energy associated with tunneling,
which is an excellent approximation for the lattice depths explored
in this work.  For $U = 50$ $E_R$, the central lattice site contains
roughly 120 atoms, with a $1/e$ radius of 19 lattice sites, where $g
\beta/\hbar \sim 17$ rad/s and $\gamma/\hbar \sim$ 0.6 rad/s.

\begin{figure}[htb!]
\includegraphics[scale = .49, bb = 50 50 450 500]{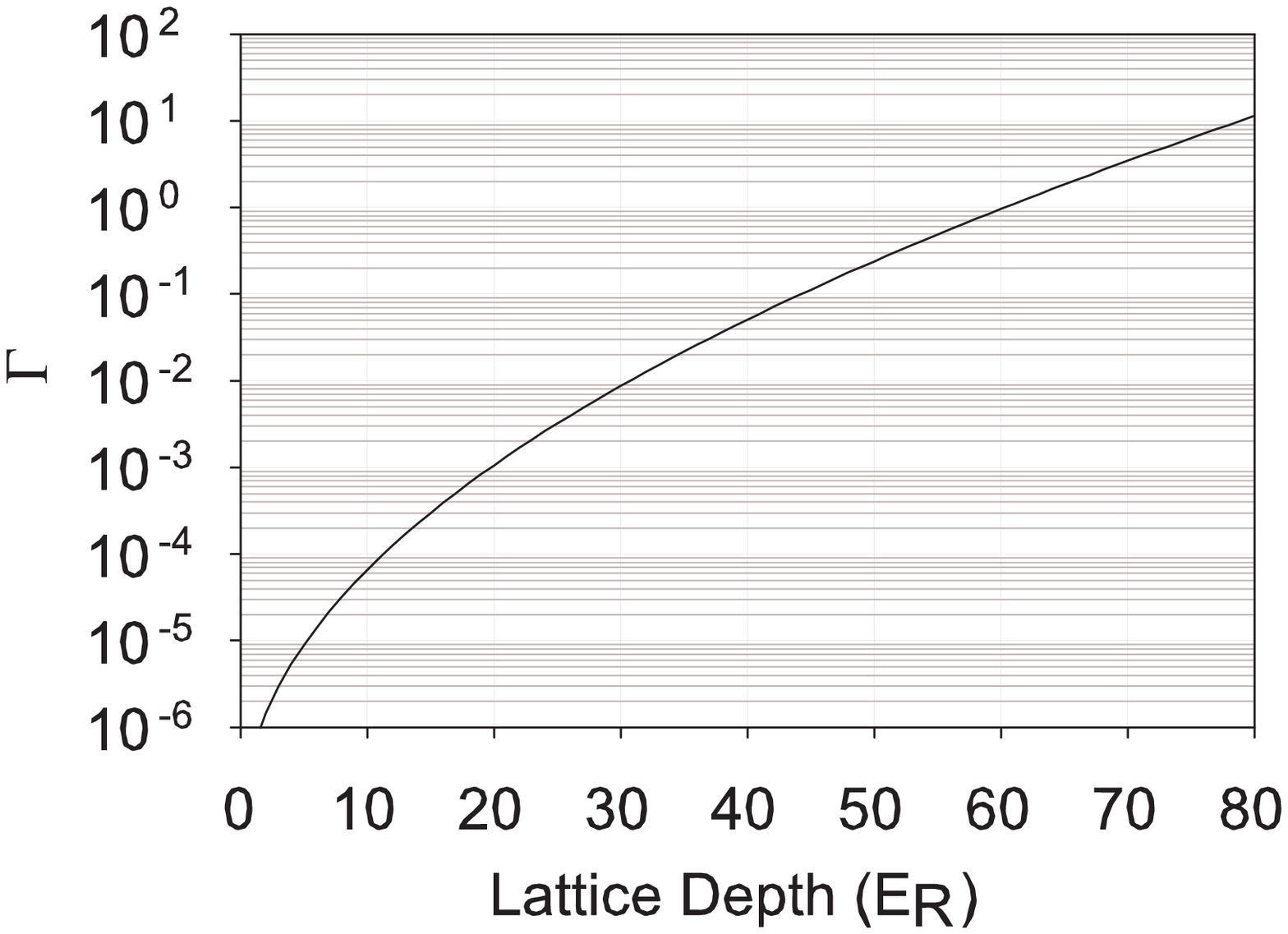} \caption {\newline {\protect\small
$\Gamma \equiv E_C/E_J \equiv g\beta/N\gamma$ is calculated as a
function of lattice depth.  For a translationally invariant lattice
system, $\Gamma \sim 1$ denotes the onset of a Mott-Insulating
regime.}} \label{Figure1}
\end{figure}

Accurate determination of $\Gamma$ (Fig.~\ref{Figure1}) further
hinges on measurement calibration of atom number and lattice depth.
We infer the total number of atoms in the trapped condensate from an
absorptive image of the atomic array. We measure the lattice depth
in three independent ways. First, we determine the lattice depth
through direct observation of the harmonic oscillation frequency in
each well:  we measure atoms lost from the lattice via parametric
heating as a function of the modulation frequency of a small
amplitude perturbation to the position of the potential minimum.
Second, we measure the period of Kapitza-Dirac diffraction by
suddenly turning on the lattice \cite{kd}. Finally, we directly
measure $\gamma$ at weak lattice depths ($U$ $< 10$ $E_R$) by
observing the amplitude of the oscillation induced by an applied
transient phase gradient \cite{lattice}. In particular, a phase
shift (quasi-momentum) $\delta \phi = \pi/2$, induces an oscillation
with an amplitude (in lattice units) equal to
$\sqrt{2\gamma/\Omega}$ \cite{Smerzi}. These experimental techniques
determine the lattice calibration with an uncertainty of $\pm 8\%$,
which agrees with the calculated lattice depth based on measured
parameters to within 20$\%$.

Our experimental apparatus is described in detail in
Ref.~\cite{Orzel} and begins with $10^8$ laser cooled $ ^{87}$Rb
atoms loaded into a time-orbiting potential (TOP) trap. We use both
TOP and forced radio frequency evaporative cooling to generate
nearly pure condensates with $4 \pm 1 \times 10^3$ atoms, optically
pumped into the $F=2, m_F = 2$ state.  We estimate $T/T_c \sim 0.27$
before the lattice is turned on ($T$ is the temperature and $T_c$
the BEC transition temperature). After the BEC phase transition, we
adiabatically relax the magnetic trap, thereby increasing the
condensate's size and decreasing its density. This results in a
radial trapping frequency of $\omega_\perp = 70$ rad/s and $\omega_z
= 195$ rad/s, corresponding to $\Omega/\hbar = 4.2$ rad/s ($\Omega =
2.1\times 10^{-4} \, E_R$).  A retro-reflected laser beam in the
vertical direction at a wavelength $\lambda = 840$ nm is
superimposed over the condensed atom cloud to create a 1-D lattice
potential.  The light is focussed to a $1/e$ intensity radius of $50
\,\mu$m and is far-detuned to the red from the 780 nm optical
resonance of $ ^{87}$Rb.  This creates a standing wave which forms a
periodic array of potential wells spaced by $\lambda/2$. Atoms
occupy lattice sites at the anti-nodes of the sinusoidally varying
optical potential. The lattice also provides additional transverse
confinement of the atoms. For example, at $U= 50$ $E_R$, the
characteristic trap frequencies are 842 rad/s (transverse) and 3.1
$\times$ 10$^5$ rad/s (longitudinal).

We slowly increase the intensity of the lattice laser (in a time of
200 ms) to minimize non-adiabaticity in the state preparation. We
note that this ramp may not be strictly adiabatic for the deeper
lattice depths used in this work \cite{Juha,Ruos2}. In order to
study the transport properties of atoms through the array, we
suddenly shift the harmonic potential minimum and stroboscopically
observe the ensuing motion of the center-of-mass of the array (using
destructive absorptive imaging techniques). We shift the harmonic
potential by pulsing on a weak vertical magnetic field. The
resulting translation, $\Delta$, to the minimum induces a chemical
potential offset, $E = 4\Delta\Omega/\lambda$, between adjacent
wells, which drives subsequent array dynamics.

\begin{figure}[htb!]
\includegraphics[scale = 1.43]{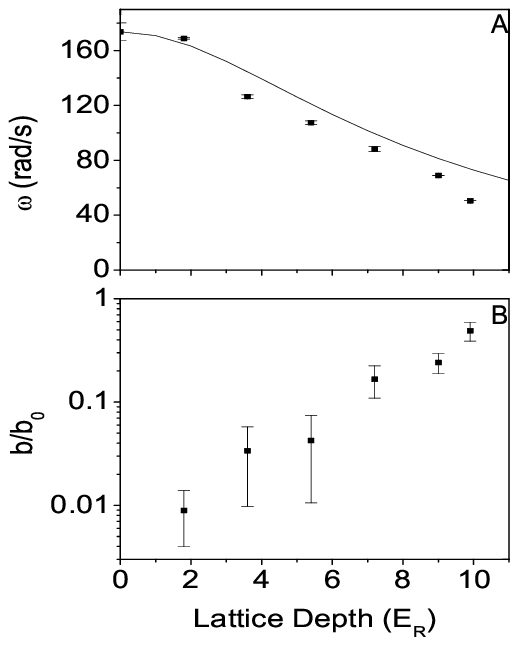} \caption {\newline {\protect\small (A) Center-of-mass
oscillation frequency vs. lattice depth for $\Delta = $ 3.5 $\mu$m.
Solid line is solution to semiclassical hydrodynamic equations. (B)
Damping of center-of-mass oscillation vs. lattice depth. $b/b_0$
notation is adopted for comparison with Ref.~\cite{Fertig}, where
$b_0\equiv 2m\omega_0$, and $\omega_0$ is the magnetic harmonic trap
frequency.}} \label{Figure2}
\end{figure}

We investigate three regimes of lattice transport: $\Gamma \ll 1$,
$\Gamma < 1$ and $\Gamma \geq 1$.  In order to tie in with previous
work \cite{Ingu,Fertig}, we first explore the superfluid regime for
low lattice depths ($\Gamma \ll 1$) where the semiclassical
hydrodynamic equations are well suited. Transport oscillation
frequencies closely follow a scaled harmonic magnetic trap
frequency, $\omega=\omega_0\sqrt{m/m^*}$ with an effective mass, $
m/m^* = \gamma\pi^2/2E_R$, determined by the lattice depth
\cite{Stringari}. We characterize the observed oscillations through
their amplitude and frequency, determined by non-linear least
squares fits to the oscillation time sequences
[$z(t)=Ae^{-tb/2m^*}\cos(\omega t)$, where $A$ denotes the
oscillation amplitude and $b$ the damping coefficient]. In
Fig.~\ref{Figure2}A, we see good correlation of our observed
frequencies with semiclassical theory. Ground state coherent
transport in this regime is also contingent on the trap displacement
being sufficiently small so that an unstable regime is avoided.

\begin{figure}[htb!]
\includegraphics[scale = .49,bb = 50 180 450 600]{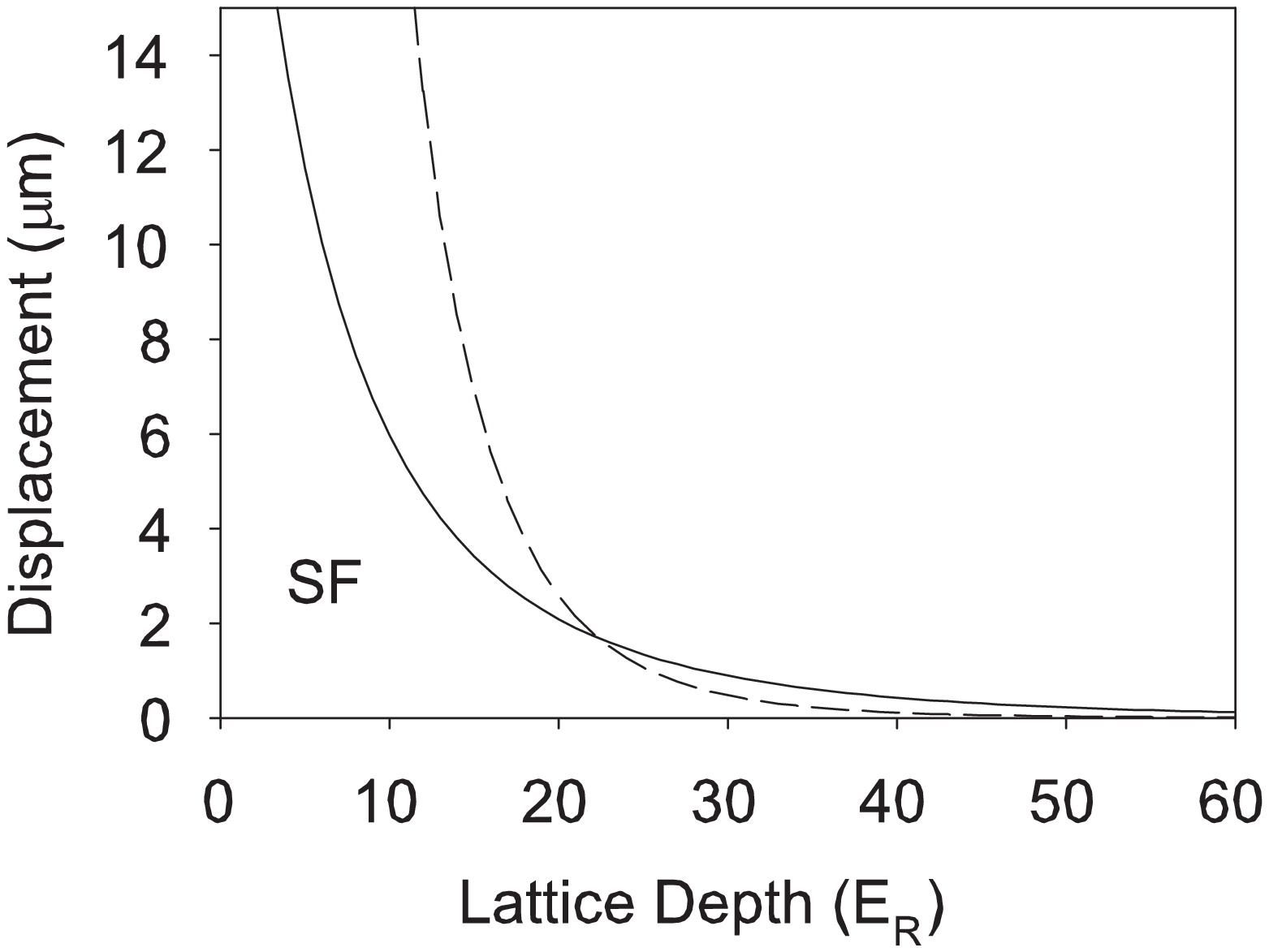} \caption {\newline {\protect\small
Critical displacement, $\Delta_{\mathrm{crit}}$, as predicted by the
GPE, is plotted as a function of lattice depth, shown with a solid
line. Above the solid line this semiclassical picture predicts a
dynamically unstable, dissipative regime, and below the line,
superfluid transport (SF) is expected.  The dashed line indicates
the condition where $\gamma = E$. For $E > \gamma$ and weak
interactions, lattice dynamics are governed by Bloch oscillations.
}} \label{Figure3}
\end{figure}

If the trap displacement is too large, the system enters a
dynamically unstable regime, where small perturbations around a
plane wave grow exponentially in time \cite{Ingu2}. This is expected
to occur after a critical displacement, $\Delta_{\mathrm{crit}} =
\lambda/2\sqrt{2\gamma/\Omega}$, where a $\pi/2$ phase difference
accumulates between adjacent wells \cite{Smerzi}. In this regime,
dissipation is expected to lead to rapid frustration of coherent
tunneling. For $\Delta = 6.4\,\mu$m, the largest displacement used
in this work, this occurs for $U> 10$ $E_R$ (see
Fig.~\ref{Figure3}).

The data shown in Fig.~\ref{Figure2}, however, demonstrates
significant damping, despite being taken with $\Delta <
\Delta_{\mathrm{crit}}$. Furthermore, for this data $\gamma \gg E$,
indicating that the width of the lowest band is greater than the
energy offset induced between wells (see Fig.~\ref{Figure3}). Thus,
we also do not expect localization related to Bloch oscillations.
One possible explanation for the damping seen in Fig.~\ref{Figure2}B
is effects due to quantum fluctuations, predicted to occur even at
low lattice depths \cite{Polkovnikov, Ruos}. This effect has
recently been suggested as a potential mechanism for experimentally
observed damping in Ref.~\cite{Fertig}.  We note that the damping
observed in Ref.~\cite{Fertig} is significantly greater than what we
observe, however, in that system, lattice site occupation is on the
order of unity.  Other possible explanations for our observed
damping include finite temperature effects \cite{Ingu3}.

\begin{figure}[htb!]
\includegraphics[scale=1.76]{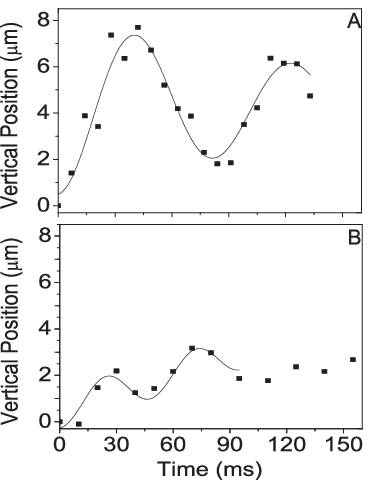} \caption {\newline {\protect\small
Center-of-mass oscillations with $\Delta = 3.5\,\mu$m are shown for
(A) the underdamped regime, with $\omega = 88.2 \pm 1.9$ rad/s for
$U=7.2 \, E_R$ and (B) the overdamped regime, with $\omega =
149.9\pm 5.8$ rad/s, taken at $U = 10.8 \, E_R$.}} \label{Figure4}
\end{figure}

\begin{figure}[htb!]
\includegraphics[scale = .52,bb = 120 110 500 700]{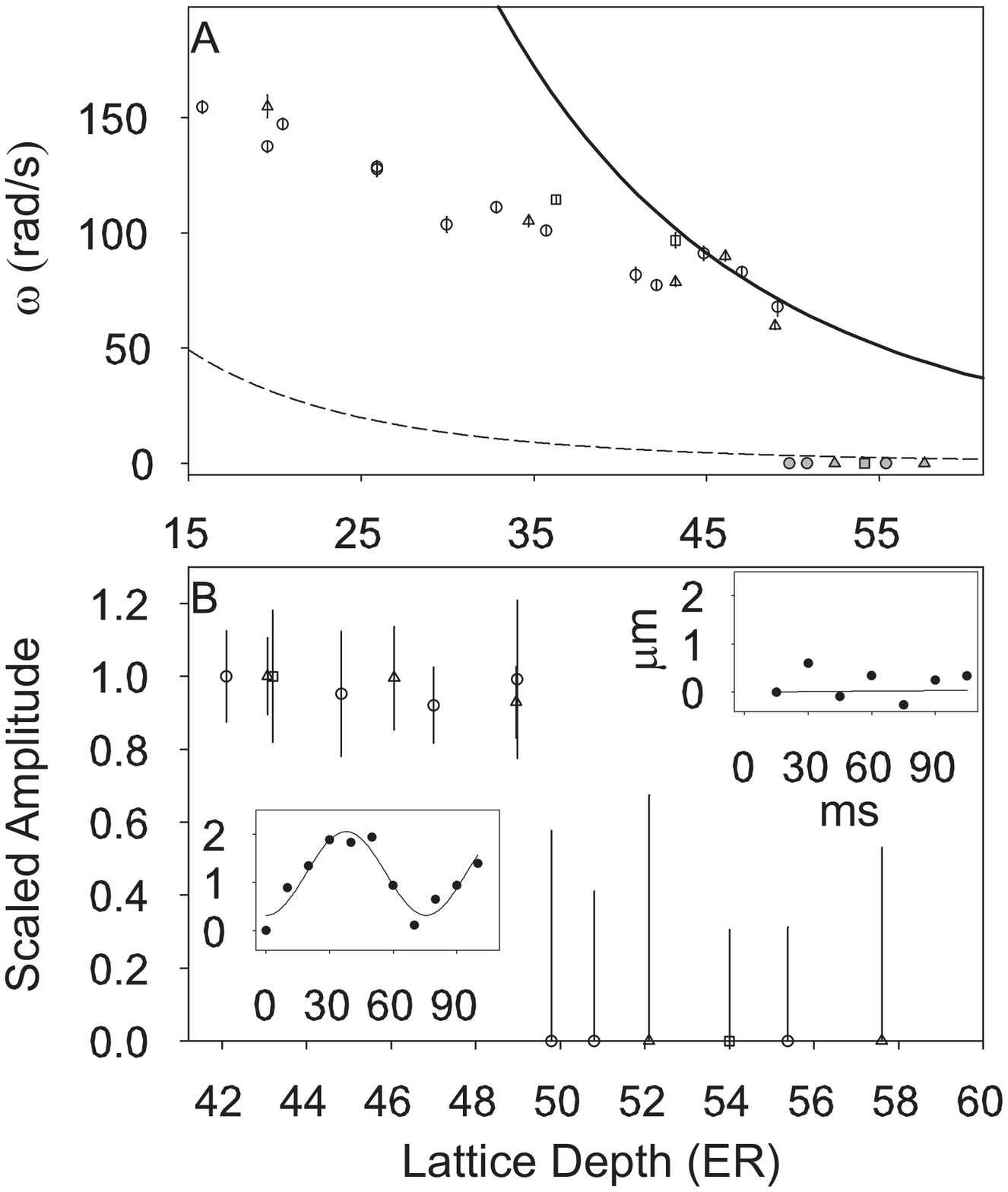} \caption {\newline {\protect\small
(A) Center-of-mass oscillation frequency, $\omega$, vs. lattice
depth for harmonic trap displacements $\Delta = $ 6.4 $\mu$m
(circles), 4.3 $\mu$m (triangles), and 3.2 $\mu$m (squares). The
solid line indicates the excitation frequencies for a lattice array
phonon mode with an effective length of 4 sites, and the dashed line
denotes the oscillation frequency predicted by the semiclassical
hydrodynamic equations.  Filled points, shown with $\omega =0$,
represent data where we cannot experimentally resolve an
oscillation. (B) Center-of-mass oscillation amplitude vs. lattice
depth. Typical oscillation sequences are shown in the upper and
lower insets, for parameters $U = 54 \, E_R$, $\Delta = 6.4$ $\mu$m
and $U = 47 \, E_R$, $\Delta = 6.4$ $\mu$m respectively. Below the
transition, errors are inferred from the residuals of the non-linear
least squares fits. Above the transition, error bars are determined
from the variance of the ensemble of data points. Amplitudes are
scaled to the measured oscillation amplitude for $U= 42\,E_R$.}}
\label{Figure5}
\end{figure}

For $\Gamma < 1$ we observe higher frequency oscillations which
emerge on top of the overdamped slower oscillations.
Fig.~\ref{Figure4}A shows a slow oscillation in the underdamped
region with $\omega = 88.2 \pm 1.9$ rad/s at $U=7.2 \, E_R$.
Fig.~\ref{Figure4}B displays the emergence of a faster oscillation
with $\omega = 149.9\pm 5.8$ rad/s, taken at $U = 10.8 \, E_R$,
where the non-linear fit requires an added linear term to account
for the initial slope. We continue to observe high frequency
transport oscillations as the lattice is increased for $U<50$ $E_R$,
as shown in Fig.~\ref{Figure5}A.  We rule out the possibility of
these high frequency oscillations being Bloch oscillations since
$\omega$ is independent of $\Delta$, which determines the energy
offset between adjacent wells.

The presence of these oscillations cannot be explained within the
ground state semiclassical picture, since the observed frequencies
are much greater than those expected from the hydrodynamic equations
(see dashed line in Fig.~\ref{Figure5}A) \cite{Stringari}. In
addition, our observed frequencies cannot be explained by using
standard approximations to include the role of quantum fluctuations
in ground state transport. Recent theoretical approaches using the
Truncated Wigner approximation (TWA) have shown that quantum
fluctuations tend to reduce transport frequencies in the overdamped
regime and to increase damping \cite{Polkovnikov}.

We expect, however, that the observed frequencies in this regime
should be related to the characteristic frequencies for phonon
excitations.  An array phonon frequency can be determined for an
effective phonon wave-vector, $q_\mathrm{eff}$, which characterizes
the array coherence length. We infer $q_\mathrm{eff}$ from the
phonon dispersion relation for a uniform lattice $ \hbar\omega_q =
\sqrt{4\gamma \sin^2 (\frac{q\lambda}{4})[2N_ig\beta+4\gamma
\sin^2(\frac{q\lambda}{4})]}$ \cite{Burnett}.  We find that for
suitably deep lattice depths (near the expected MI cross-over), our
observed frequencies correspond to phonon excitations with $4
\pi/q_\mathrm{eff} \lambda = 4$ lattice sites, as seen in
Fig.~\ref{Figure5}A.  It is interesting to note that this phonon
frequency also corresponds to the generalized Josephson frequency
associated with a 2-well system. As described earlier, the
complexity of the Hilbert space of our system makes achieving a
quantitative theoretical prediction for $q_\mathrm{eff}$ very
difficult. However, our empirical observation of an effective phonon
excitation length (4 sites), much less than the full spatial extent
of the atomic cloud (19 sites), may be due to the emergence of MOT
domains in the outer wells \cite{Ketterle,Gerbier}.  For shallower
lattice depths (in the regime shown in Fig.~\ref{Figure5}A) we
observe frequencies which are consistent with longer range phonon
modes, possibly due the effective array coherence length increasing
with reduced $\Gamma$ \cite{Zwerger}.

We note that for these lattice depths, our experiments probe the
ill-understood regime where the semiclassical
(Gross-Pitaevskii/hydrodynamic) equations become dynamically
unstable (see Fig.~\ref{Figure3}). In this regime, for example, the
Bogoliubov dispersion relations have imaginary solutions and
unconventional normalization \cite{Fetter}, and thus, the TWA is
unreliable. It has been shown that the exact quantum dynamics of a
dynamically unstable system diverge (logarithmically with $N$) from
the classical trajectories \cite{Castin, Anglin}, which further
complicates the analysis of our excited state transport.

For $\Gamma \sim 1$ we observe an abrupt cessation of superfluid
transport.  The oscillation amplitude falls to zero at a critical
value for the lattice depth corresponding to $\Gamma \sim 1$
(Fig.~\ref{Figure5}B), consistent with the expected superfluid-MI
transition \cite{Doniach,Monien} (at $ U = 49$ $E_R$, $\Gamma =
0.21$ with an error range of $0.09<\Gamma <0.52$ determined,
predominantly, by our experimental uncertainty in lattice
calibration). Fig.~\ref{Figure5}B (see lower inset) illustrates the
oscillation induced by a shift ($\Delta =$ 6.4 $\mu$m) in the
harmonic potential at a lattice depth of $U = $ 47 $E_R$ and is
contrasted with localization observed for the same displacement at
$U = $ 54 $E_R$ (see upper inset). The observed threshold is
independent of the lattice displacement (within our experimental
limits). Additionally, above the transition, the atoms are observed
to localize immediately after the shift in position of the harmonic
potential, as illustrated in the inset to Fig.~\ref{Figure5}B. These
observations explicitly rule out other localization mechanisms --
such as macroscopic quantum self-trapping \cite{MQST,Leggett} or
dynamical instabilities \cite{Smerzi} -- where the localization
depends on the initial displacement, or manifests itself only after
the array dynamically evolves following the displacement (in the
case of dynamical instabilities).

In conclusion, we have used the presence of anomalous phonon
excitations to probe atom transport deep in the dissipative regime.
This regime has been previously inaccessible using ground state
array transport. We observe localization to occur at the critical
value $\Gamma \sim 1$, which suggests localization due to quantum
fluctuation effects. This work raises the possibility for further
control of macroscopic quantum coherence, even in excited state
systems.

This work was supported by the National Science Foundation, the Army
Research Office and the Office of Naval Research.  S. D.
acknowledges support from the German Academic Exchange Service
(DAAD).  We thank A. Fetter, T.L. Ho, S. Sachdev, S. Girvin, and A.
Polkovnikov for useful discussions and C. Orzel and M. Fenselau for
experimental contributions during the early stages of this work.


\begin{thebibliography}{}
\bibitem{lattice} We apply the transient phase gradient by
suddenly switching the magnitude of the rotating component,
$B_{rot}$, of the TOP magnetic field. Since the curvature of a TOP
trap scales as ${B_{q}^2}/{B_{rot}}$, a sudden increase in $B_{rot}$
weakens the magnetic trap and the atoms are exposed to a
gravity-induced potential gradient along the vertical array.  For
large enough values of $B_{rot}$, gravity imprints a potential
energy difference $mg{\lambda}/{2}$ between adjacent wells for the
duration $\tau$ of the pulse.  If the phase in each well is defined,
it precesses as $\phi_j = \phi_{j_0} + mg({j\lambda}/{2})\tau$ where
all initial phases $\phi_{j_0}$ are equal.
\end{thebibliography}
\end{document}